\documentclass[12pt,a4paper]{article}
\usepackage{amssymb,amsmath,verbatim,graphicx}

\newtheorem{definition}{Definition}
\newtheorem{theorem}{Theorem}
\newtheorem{example}{Example}

\def\var{\mathop{\rm var}\nolimits}%
\def\cov{\mathop{\rm cov}\nolimits}%
\def\Normal{\mathop{\rm Normal}\nolimits}%

\begin{document}

\title{Principle of Detailed Balance and Convergence Assessment of Markov Chain Monte Carlo methods and Simulated Annealing}
\author{Ioana A. Cosma and Masoud Asgharian\footnote{Ioana A. Cosma is a doctoral student in the Department of Statistics, University of Oxford, 1 South Parks Road, Oxford, OX1 3TG, United Kingdom (email: cosma@stats.ox.ac.uk); Masoud Asgharian is Associate Professor, Department of Mathematics and Statistics, McGill University, Burnside Hall, 805 Sherbrooke W., Montreal, Quebec, Canada, H3A 2K6 (email: masoud@math.mcgill.ca).  This research was partially supported by research grants from NSERC and FQRNT.  The authors thank Russell Steele for insightful discussions on the topic.}}

\maketitle

\begin{abstract}
Markov Chain Monte Carlo (MCMC) methods are employed to sample from a given distribution of interest, $\pi$, whenever either $\pi$ does not exist in closed form, or, if it does, no efficient method to simulate an independent sample from it is available.  Although a wealth of diagnostic tools for convergence assessment of MCMC methods have been proposed in the last two decades, the search for a dependable and easy to implement tool is ongoing. We present in this article a criterion based on the principle of detailed balance which provides a qualitative assessment of the convergence of a given chain.  The criterion is based on the behaviour of a one-dimensional statistic, whose asymptotic distribution under the assumption of stationarity is derived; our results apply under weak conditions and have the advantage of being completely intuitive.  We implement this criterion as a stopping rule for simulated annealing in the problem of finding maximum likelihood estimators for parameters of a 20-component mixture model. We also apply it to the problem of sampling from a 10-dimensional funnel distribution via slice sampling and the Metropolis-Hastings algorithm.  Furthermore, based on this convergence criterion we define a measure of efficiency of one algorithm versus another.

\vspace{.05 in}
\end{abstract}  

\begin{center}
{1. INTRODUCTION}
\end{center}
Let $\pi$ be a given distribution such that either $\pi$ does not exist in closed form or no efficient method to simulate an independent sample from it is available.  Suppose that interest lies in the expected value of a random variable $h(X)$, denoted by $\mathbb{E}_{\pi}\big(h(X)\big)$, where $X$ has distribution $\pi$.  Monte Carlo sampling methods (Hammersley and Handscomb 1964) such as rejection sampling, importance sampling or sampling-importance resampling (SIR) approximate the value of $\mathbb{E}_{\pi}\big(h(X)\big)$ by sampling from a distribution $g$ that closely resembles $\pi$ (Smith and Gelfand 1992).  Although for low dimensional distributions $\pi$ it is oftentimes possible to find sampling distributions $g$ that provide estimates to within given accuracy with low computational cost, these sampling methods suffer greatly from the curse of dimensionality.   

The need to approximate the value of high dimensional integrals arising in statistical mechanics led to the development of MCMC sampling methods.  The first MCMC method, known today as the Metropolis Monte Carlo algorithm, was proposed by Metropolis, Rosenbluth, Rosenbluth, Teller, and Teller (1953) as a general method for studying the equilibrium properties of systems consisting of many interacting particles.  The algorithm simulates the behaviour of the system under equilibrium, and the expected value of a given property is approximated by ergodic averages based on these simulations.  In statistical terms, the Metropolis Monte Carlo algorithm constructs an ergodic Markov chain $\{X_t, t=1,\ldots,n \}$ with stationary distribution $\pi$, i.e.\ as  the number of iterations $n$ tends to $\infty$, the conditional distribution of $X_n$ given the value of $X_1$ converges to $\pi$ regardless of the starting distribution $g$, where $X_1$ has distirubtion $g$ (in notation: $X_1 \sim g$).      

Hastings (1970) generalized the procedure of proposing the next move $X_t$ given $X_{t-1}=x_{t-1}$.  His algorithm, known as the Metropolis-Hastings algorithm, transforms an arbitrary stochastic matrix into a $\pi$-reversible one, and only requires that $\pi$ be known up to a normalizing constant.  An equally popular MCMC algorithm is the Gibbs sampler, introduced by Geman and Geman (1984) with an application to image restoration.  This algorithm proposes the next move by sampling from the full conditional distributions and, unlike the Metropolis-Hastings algorithm, accepts each proposal with probability 1.  Two well-known variants on Gibbs sampling are the data-augmentation algorithm of Tanner and Wong (1987) and the substitution sampling algorithm of Gelfand and Smith (1990).  

The goal of MCMC methods is to produce an approximate i.i.d.\ sample \\
$\big\{X_{K+1}, X_{K+2},\ldots,X_{K+n} \big\}$ from $\pi$, where $K$, $n>1$, and $K$ is known as the number of `burn-in' iterations to be removed from the beginning of the chain.  Analysing the output of an MCMC method consists of assessing convergence to sampling from $\pi$, convergence to i.i.d.\ sampling, and convergence of empirical averages of the form $\frac{1}{n} \sum_{i=1}^{n} h(X_{K + i})$ to $\mathbb{E}_{\pi} \big( h(X) \big) = \int \ h(x)\pi(x)dx$ as $n \to \infty$.  Robert and Casella (2004) argue that while convergence to $\pi$ is not of major concern since it can only be achieved asymptotically, the issues of convergence to i.i.d.\ sampling and of convergence of empirical averages are strongly interrelated and depend on the mixing speed of the chain.  By definition, a chain whose elements converge rapidly to weakly correlated draws from the stationary distribution is said to possess good mixing speed.  Therefore, the mixing speed of a chain is determined by the degree to which the chain escapes the influence of the starting distribution and by the extent to which it explores the high density regions of the support of $\pi$.                         

Recent research in MCMC methodology has focused on developing, on one hand, samplers that escape quickly the attraction of the starting distribution as well as that of local modes, and, on the other hand, convergence assessment criteria for analysing the mixing speed of a given chain.  A recent sampling algorithm which exploits the idea of jumping between states of similar energy to facilitate efficient sampling is the equi-energy sampler of Kou \textit{et al.}(2006).  Robert (1995,1998), Cowles and Carlin (1996), and Brooks and Roberts (1998) present a comprehensive review of the practical implementation of convergence criteria and the mathematics underlying them.  Liu (2001), Neal (1993), Brooks (1998), and Kass, Carlin, Gelman, and Neal (1998) offer an in-depth introduction to MCMC methodology and its applications, as well as discussions on the issues surrounding it. 

The common view among researchers and practitioners is that developing a good sampler or a reliable convergence criterion is problem-specific.  A sampler with good mixing speed when sampling from a relatively smooth, low-dimensional distribution might become trapped in a well of low probability when sampling from a distribution having many local modes.  Similarly, a convergence criterion which proves reliable for analysing a given MCMC output might incorrectly assess the convergence of a chain that has only explored a subset of the entire support space. Our interest lies in convergence assessment, in particular, in identifying lack of convergence.  We define a one-dimensional statistic and derive an intuitive criterion based on the principle of detailed balance that provides a qualitative assessment on the convergence of a given MCMC chain.

In Section 2 we recall basic notions and results from the theory of Markov chains, which we subsequently use in Section 3 to derive the asymptotic distribution of our proposed statistic under the assumption of stationarity.  In the same section, we discuss two possible implementations of our criterion, one using the asymptotic distribution, the other experimental as a qualitative tool.  Section 4 discusses two applications: one as a stopping rule for simulated annealing, an algorithm for function maximization applied to the problem of finding maximum likelihood estimators (Azencott 1992), the second as a graphical tool for comparing the performances of Metropolis-Hastings versus slice sampling for the problem of sampling from a 10-dimensional funnel distribution.  All computations were performed using code written in C++.  We conclude in Section 5 with general remarks, comparisons, and criticisms.  

\begin{center}    
2. PRELIMINARIES
\end{center}
\noindent
Let $X = \{X_t, \ t=1,2,\ldots \}$ be a Markov chain with state space $S$ and transition probability matrix $P=(p_{ij})$.  We refer the reader to Medhi (1994), Norris (1997), and Jones (2004) for details and proofs.  For the purpose of the convergence criterion we present in this article, we restrict our attention to finite Markov chains.  

Let $p_{ij}^{(n)}$ be the transition probability from state $i$ to state $j$ in $n$ steps.  The Ergodic Theorem states that if $X$ is irreducible and aperiodic, then the limits $\pi_j := \lim_{n \to \infty}p_{ij}^{(n)}$ exist and are independent of the initial state $i$ for all $i,j \in S$ and $( \pi_j, j \in S )$ is the stationary distribution of $X$.  The chain $X$ is called ergodic.

\begin{definition} ({\it Principle of detailed balance})
Transition probability matrix $P$ and probability distribution $\pi$ are said to be
in \textit{detailed balance}, or, equivalently, \textit{the principle of detailed balance} is said to hold, if
\begin{math} 
\pi_i p_{ij} = \pi_j p_{ji} \ \forall i,j \in S.
\end{math}
\end{definition}

\begin{definition} 
A Markov chain $X$ with irreducible transition probability matrix $P$ and initial distribution $g$, i.e.\ $X_1 \sim g$, is \textit{reversible} if, for all $N \geq 2$, the chain $\{X_N, X_{N-1}, \ldots, X_2, X_1 \}$ is a Markov chain with transition probability matrix $P$ and initial
distribution $g$.  
\end{definition}

Norris (1997) proves that if $X$ is irreducible, then it is reversible if and only if $P$ and $g$ are in detailed balance, where $g$ is the initial distribution of $X$.  The following definitions are needed to introduce the Markov chain Central Limit Theorem (Jones 2004).

\begin{definition} \label{def:bounds}
Let $M(i)$ be a nonnegative function and $\gamma(n)$ a nonnegative decreasing function on the positive integers such that
\begin{equation}\label{condition}
\parallel P^n(i,\cdot) - \pi(\cdot) \parallel \leq M(i)\gamma(n).
\end{equation}
Let $X$ be a Markov chain on state space $S$ with transition probability $P$ and stationary distribution $\pi$.  If (\ref{condition}) holds for all $i \in S$ with $\gamma(n) = t^n$ for some $t < 1$, then $X$ is geometrically ergodic.  If, moreover, $M$ is bounded, then $X$ is uniformly ergodic.  If (\ref{condition}) holds for all $i \in S$ with $\gamma(n) = n^{-m}$ for some $m \geq 0$, then $X$ is polynomially ergodic of order $m$.
\end{definition}

\begin{theorem}{\textbf{The Central Limit Theorem} (finite state space)} \label{CLT}
Let $X$ be an ergodic Markov chain on state space $S$ with stationary distribution $\pi$.  Let $h: S \to \mathbb{R}$ be a Borel function.  Assume that one of the following conditions holds:
\begin{enumerate}
\item $X$ is polynomially ergodic of order $m > 1$, $E_{\pi}M < \infty$ and there exists $B < \infty$ such that $|h(X)| < B$ almost surely; 
\item $X$ is polynomially ergodic of order $m$, $E_{\pi}M < \infty$ and $E_{\pi} \big( |h(X)|^{2 + \delta} \big) < \infty$ where $m\delta > 2 + \delta$; 
\item $X$ is geometrically ergodic and $E_{\pi} \big( |h(X)|^{2 + \delta} \big) < \infty$ for some $\delta > 0$;
\item $X$ is geometrically ergodic and $E_{\pi} \big (h^2(X)[\log^{+}|h(X)|] \big ) < \infty$; 
\item $X$ is geometrically ergodic, satisfies detailed balance and $E_{\pi}h^2(X) < \infty$; \item $X$ is uniformly ergodic and $E_{\pi} \big (h^2(X) \big) < \infty$.
\end{enumerate}
Then for any initial distribution,
\begin{displaymath}
\sqrt{n}\Big (\bar{h}_n - E_{\pi} \big(h(X) \big) \Big) \stackrel{\mathcal{D}}{\to} \Normal \big(0,\sigma^2_h \big) \ \textit{as $n \to \infty$,}
\end{displaymath}
where $\bar{h}_n = \frac{1}{n}\sum_{i=1}^n h(X_i)$ and
\begin{math}
\sigma^2_h = \var_{\pi}\big (h(X_1) \big) + 2\sum_{i=2}^{\infty} \cov_{\pi}\big (h(X_1),h(X_i) \big ) < \infty.
\end{math}
\end{theorem}

\begin{center}
3. DETAILED BALANCE AND CONVERGENCE DIAGNOSTICS
\end{center}

Let $\pi=(\pi_i, \ i \in S)$ be a discrete distribution with finite state space $S$, $m=|S|$.  Let $\{X_t, \ t=1,\ldots,n\}$ be an irreducible, aperiodic Markov chain with transition probability matrix $P=(p_{ij})$ and stationary distribution $\pi$.  We say that a chain has reached equilibrium by step $t$ if $P^{t}(i,j)=\pi_j$, $\forall i,j \in S$ and $\exists i,j \in S$ such that $P^{t-1}(i,j) \neq \pi_j$.  Our convergence assessment criterion is based on the principle of detailed balance from statistical mechanics (Chandler 1987).  Statistical mechanics is concerned with the study of physical properties of systems consisting of very large number of particles, for example liquids or gases, as these systems approach the equilibrium state, i.e.\ a uniform, time-independent state.  In these terms, the principle of detailed balance states that a physical system in equilibrium satisfies
\begin{displaymath}
\frac{\pi_i}{\pi_j} = \frac{p_{ji}}{p_{ij}} = \exp \Big (-\frac{E_i-E_j}{kT} \Big ), \ \forall i,j \in S,
\end{displaymath}
where $E_i$ is the energy of the system in state $i$, $k$ is Boltzmann's constant, $T$ is the temperature, and $\pi_i$ and $p_{ij}$ have the usual interpretation. 

We assume that the Markov chain $\{X_t, \ t=1,\ldots,n\}$ is constructed to satisfy detailed balance.  This is oftentimes the case since the principle of detailed balance implies that $\pi$ is the stationary distribution of the chain, and it is easier to check the former than the latter, see for example the discussions on the Metropolis-Hastings (Hastings 1970) and slice sampling algorithms (Neal 2003).  We introduce the notion of an energy function $E_i \propto -\log(\pi_i)$, $\forall i \in S$.  When implementing simulated annealing, the stationary distribution at temperature $T_k$ is $\pi^{1/T_k}$, so the energy function becomes $E_i=-\log(\pi_i)/T_k$, where $\{T_k, \ k=1,2,\ldots \}$ is a sequence of decreasing temperatures.  Therefore, the equilibrium probability of being in state $i$ equals
\begin{math}
\pi_i = \frac{1}{Z} \exp(-E_i),
\end{math}
where the normalizing constant is defined as $Z:=\sum_{i \in S} \exp(-E_i)$.  Define the following approximation to $\pi_i$ based on a Markov chain of $n$ iterations
\begin{displaymath}
\hat{\pi}_i = \frac{1}{n} \sum_{j=1}^n \mathbb{I}(X_j=i), \ \forall i \in S.
\end{displaymath}

\indent
The idea of working with indicator functions is similar to that of Raftery and Lewis (1992) who develop a convergence assessment method based on the sequence $\{ \mathbb{I}(X_t \leq i), \ t=1,\ldots \}$, for fixed $i \in S$.  We point out that, for fixed $i \in S$, the sequence $\{ \mathbb{I}(X_t=i), \ t=1,\ldots \}$ forms a Markov chain, whereas the sequence defined by Raftery and Lewis does not.  Brooks \textit{et al.}(2003) use a similar approach of estimating the stationary distribution by the empirical distribution function obtained from the MCMC output; they derive nonparametric convergence assessment criteria for MCMC model selection by monitoring the distance, as the number of simulations increases, between the empirical mass functions obtained from multiple independent chains. 
  
\indent
Our criterion assesses the convergence of the chain by comparing the behaviour of the functions
\begin{math}
f_i=\hat{\pi}_i / \exp(-E_i), \ i \in S, \ \textrm{to their average $\bar{f}=\frac{1}{m} \sum_{j \in S} f_j$,}
\end{math}
via the statistic
\begin{math}
V_n := \frac{n}{m} \sum_{i \in S} \big (f_i - \bar{f} \big)^2.
\end{math} 

\begin{flushleft}
3.1 Theoretical approach
\end{flushleft}

We proceed to derive the distribution of the statistic $V_n$ under the hypothesis that the chain has reached stationarity, i.e.\ that $X_i \sim \pi, \ \forall i=1,\ldots,n$.
\begin{displaymath}
V_n =  \frac{n}{m} \sum_{i \in S} \bigg \{ f_i - \frac{1}{m} \sum_{j \in S} f_j \bigg \}^2 
 =  \frac{n}{m} \sum_{i \in S} \bigg \{ f_i - \frac{1}{m} f_i - \frac{1}{m} \sum_{\stackrel{j \in S}{j \neq i}} f_j \bigg \}^2 
 =  \frac{n}{m} \sum_{i \in S} \big \{ \mathbf{a_i}' \mathbf{f} \big \}^2,
\end{displaymath}
where $\mathbf{f}=(f_i, i \in S)'$ and $\mathbf{a_i}=\big(-\frac{1}{m},\ldots,-\frac{1}{m},1-\frac{1}{m},-\frac{1}{m},\ldots,-\frac{1}{m} \big)'$ is an $m$-dimensional column vector with $i$th entry equal to $1-\frac{1}{m}$ and the remaining entries equal to $-\frac{1}{m}$.  Define the following $(m \times m)$ dimensional matrix
\begin{displaymath}
\mathbf{A} = \left( \begin{array}{c}
\mathbf{a_1}' \\
\mathbf{a_2}' \\
\vdots \\
\vdots \\
\mathbf{a_m}'
\end{array} \right )
= \left( \begin{array}{ccccc}
1-\frac{1}{m} & -\frac{1}{m} & -\frac{1}{m} & \ldots & -\frac{1}{m} \\
-\frac{1}{m} & 1-\frac{1}{m} & -\frac{1}{m} & \ldots & \vdots \\
\vdots & \vdots & \ddots & \vdots & \vdots \\
-\frac{1}{m} & \ldots & \ldots & 1-\frac{1}{m} & -\frac{1}{m} \\
-\frac{1}{m} & \ldots & \ldots & -\frac{1}{m} & 1-\frac{1}{m}
\end{array} \right),
\end{displaymath}
so $V_n = \frac{n}{m}\big \{ \mathbf{A}\mathbf{f} \big \}' \big\{ \mathbf{A}\mathbf{f} \big \}$.

\noindent
First, we observe that $\forall i \in S$,
\begin{equation} \label{derivation:1}
\big (f_j - \mathbb{E}_{\pi}f_j, \ j \in S \big) \mathbf{a_i}' =  \Big (1-\frac{1}{m}\Big) \Big [f_i - \mathbb{E}_{\pi} f_i \Big ] - \frac{1}{m} \sum_{\stackrel{j \in S}{j \neq i}} \Big (f_j - \mathbb{E}_{\pi} f_j \Big)  
= f_i - \bar{f},
\end{equation}
since $\mathbb{E}_{\pi} f_j = \frac{1}{Z}, \ \forall j \in S$.  
Second, we notice that
\begin{equation} \label{derivation:2}
f_i - \mathbb{E}_{\pi} f_i = \frac{\hat{\pi}_i}{e^{-E_i}} - \frac{1}{Z} = \frac{\hat{\pi}_i - \pi_i}{Z\pi_i}, \ \forall i \in S.
\end{equation}

\noindent
Define $W_{i,n} := \sqrt{n} \big ( \hat{\pi}_i - \pi_i \big ), \ \forall i \in S$, and the $m$-dimensional column vector $W_n:=\big(W_{i,n}, i \in S \big)'$.  From (\ref{derivation:1}) and (\ref{derivation:2}), we obtain that
\begin{math}
V_n = \big \{ \mathbf{C}W_n \big \}' \big \{\mathbf{C}W_n \big\},
\end{math}
where
\begin{displaymath}
\mathbf{C}=\mathbf{A} \left (\begin{array}{cccc}
\frac{1}{\sqrt{m}Z\pi_1} & 0 & \ldots & 0 \\
0 & \frac{1}{\sqrt{m}Z\pi_2} & 0 & \vdots \\
\vdots & \vdots & \ddots & \vdots \\
0 & \ldots & 0 & \frac{1}{\sqrt{m}Z\pi_m}
\end{array} \right )
= \left (\begin{array}{cccc}
\frac{m-1}{m^{3/2}e^{-E_1}} & 0 & \ldots & 0 \\
0 & \ddots & 0 & \vdots \\
\vdots & 0 & \ddots & \vdots \\
0 & \ldots & 0 & \frac{m-1}{m^{3/2}e^{-E_m}}
\end{array} \right )
\end{displaymath}

\noindent
The following result presents the asymptotic distribution of the statistic $V_n$ under the assumption of stationarity.

\begin{theorem} \label{result}
Under the conditions of Theorem~\ref{CLT}, $\mathbf{C}W_n \stackrel{\mathcal{D}}{\to} \Normal \big (\mathbf{0},\mathbf{C}\Sigma\mathbf{C}'\big)$ and  
$V_n \stackrel{\mathcal{D}}{\to} \sum_{i=1}^k \lambda_i Z_i^2$ as $n \to \infty$, where $\lambda_1,\ldots,\lambda_k$ are the characteristic roots of $\mathbf{C}\Sigma\mathbf{C}'$ and $Z_1,\ldots,Z_k$ are i.i.d.\ $\Normal(0,1)$ random variables.
\end{theorem}

\noindent{\textbf{proof:}}
We begin by pointing out that irreducible and aperiodic Markov chains on finite state spaces are uniformly ergodic (Roberts and Rosenthal 2004), so condition (6) of Theorem~\ref{CLT} is satistifed.  It follows that for every $i \in S$,
\begin{displaymath}
W_{i,n} = \sqrt{n}\big(\hat{\pi}_i - \pi_i\big) = \sqrt{n} \bigg \{ \frac{1}{n} \sum_{j=1}^n \mathbb{I}(X_j=i) - \mathbb{E}_{\pi}\big( \mathbb{I}(X_1=i) \big) \bigg \} \stackrel{\mathcal{D}}{\to} \Normal \big(0,\sigma^2_i\big) 
\end{displaymath}
as $n \to \infty$, where
\begin{displaymath}
\sigma^2_i = \pi_i(1-\pi_i) + 2\sum_{j=2}^{\infty} \Big [ P\big \{ \mathbb{I}(X_j=i)=1 | \mathbb{I}(X_1=i)=1\big \} \pi_i - \pi_i^2 \Big ] < \infty.
\end{displaymath}

\noindent
By the Cram\'er-Wold Device (Billingsley 1968, Varadarajan 1958), it follows that $W_n \stackrel{\mathcal{D}}{\to} \Normal \big(\mathbf{0},\Sigma\big)$ as $n \to \infty$, where $\mathbf{0}$ is an $m$-dimensional column vector of zeros and $\Sigma$ is an $(m \times m)$ variance-covariance matrix whose entries are given 
\begin{eqnarray*}
\Sigma(i,i) & = & \sigma^2_{i} \\
\Sigma(i,j) & = & \lim_{n \to \infty} \cov_{\pi} \big (W_{i,n},W_{j,n} \big ) = \lim_{n \to \infty} \bigg \{ \frac{1}{n}\sum_{k=1}^n \sum_{l=1}^n \cov_{\pi} \big(\mathbb{I}(X_k=i), \mathbb{I}(X_l=j) \big) \bigg \} \\
& = & \lim_{n \to \infty} \bigg \{ \frac{1}{n}\sum_{k=1}^n \Big[P \big \{X_k=i,X_k=j\big \} - \pi_i\pi_j \Big] + \frac{1}{n} \sum_{\stackrel{k,l=1}{k<l}}^n \Big[ P\big \{X_k=i,X_l=j\big \}   \\
& & - \pi_i\pi_j \Big] + \frac{1}{n} \sum_{\stackrel{k,l=1}{l<k}} \Big[ P\big \{X_k=i,X_l=j\big \} - \pi_i \pi_j \Big ]\bigg \} 
\end{eqnarray*}
So, for all $i,j \in S$, $i \neq j$
\begin{eqnarray*}
\Sigma(i,j) & = & -\pi_i\pi_j + \lim_{n \to \infty} \frac{\pi_i}{n} \bigg \{\sum_{\stackrel{k,l=1}{k<l}}^n \Big[ P\big \{X_l=j|X_k=i\big \}-\pi_j \Big] \\
& & + \sum_{\stackrel{k,l=1}{l<k}}^n \Big[ P\big \{X_l=j|X_k=i\big \}-\pi_j \Big ]\bigg \} \\
& = & -\pi_i\pi_j + 2\pi_i \sum_{k=2}^{\infty} \Big [ P\big \{X_k=j|X_1=i\big \} - \pi_j \Big ] < \infty,
\end{eqnarray*} 
The last equality follows from the fact that if a Markov chain satisfies detailed balance, then it is reversible, i.e.\ for $k>1$, $P\big \{X_{k}=j | X_1=i\big \} = P\big \{X_1=j | X_{k}=i \big \}$.  Finally, the conditions of the Markov chain Central Limit Theorem guarantee that the infinite summation in the last line is finite.

\indent
It then follows that $\mathbf{C}W_n \stackrel{\mathcal{D}}{\to} \Normal \big(\mathbf{0},\mathbf{C} \Sigma \mathbf{C}'\big)$ as $n \to \infty$.  Lastly, since $V_n=\big\{\mathbf{C}W_n\big\}' \big\{\mathbf{C}W_n\big\}$, it follows from Lemma 1 in Chernoff and Lehmann (1953) that $V_n \stackrel{\mathcal{D}}{\to} \sum_{i=1}^k \lambda_i Z_i^2$ as $n \to \infty$, where $\lambda_1,\ldots,\lambda_k$ are the characteristic roots of $\mathbf{C}\Sigma \mathbf{C}'$ and $Z_1,\ldots,Z_k$ are i.i.d.\ $\Normal(0,1)$ random variables.

\begin{flushright}
{\it Q.E.D. }
\end{flushright}

\begin{example}
Let the Markov chain be generated by the Metropolis-Hastings algorithm with symmetric proposal probability matrix $P=(p_{ij})$.  The expressions for $\Sigma(i,i)$ and $\Sigma(i,j)$ can be simplified as follows.  Consider the Markov-Bernoulli chain $\big \{ \mathbb{I}(X_j=i), \ j=1,\ldots,n  \big \}$ for fixed $i \in S$ with transition probability matrix 
\begin{math}
P_i=
\left( \begin{array}{cc}
1-a & a \\
b & 1-b 
\end{array} \right).
\end{math}
It is shown in Medhi (1994, pp. 101-102) that
\begin{displaymath}
P^{j-1}_i=
\frac{1}{a+b}  
\left( \begin{array}{cc}
b & a \\
b & a 
\end{array} \right)
+ \frac{(1-a-b)^{j-1}}{a+b}
\left( \begin{array}{cc}
a & -a \\
-b & b 
\end{array} \right), \ \forall j \geq 2.
\end{displaymath}

\noindent
Now, 
\begin{eqnarray*}
a & = &  \frac{\sum_{\stackrel{j \in S}{j \neq i}} P\big \{X_1=j, X_2=i\big \}}{1-P\{X_1 = i\}}  = \frac{P\{X_2=i\} - P\{ X_1=i, X_2=i\}}{1-P\{X_1 = i\}} = \frac{\pi_i}{1-\pi_i}\big(1-p_{ii}\big), \\
b & = & 1 - P\big \{X_2=i | X_1=i \big \} = 1-p_{ii}.
\end{eqnarray*}

\noindent
Then, provided that $\max\{0, 2\pi_i -1\} < p_{ii} < 1, \ \forall i \in S$,
\begin{eqnarray*}
\Sigma(i,i) & = & \pi_i(1-\pi_i) + 2\sum_{j=2}^{\infty} \pi_i (1-\pi_i) \Big( \frac{p_{ii}-\pi_i}{1-\pi_i} \Big)^{j-1} = \frac{\pi_i (1-\pi_i) (1+p_{ii}-2\pi_i)}{1-p_{ii}}, \\
\Sigma(i,j) & = & -\pi_i\pi_j + 2\pi_i \sum_{k=2}^{\infty} \Big (P^{k-1}(i,j) - \pi_j\Big), \quad \textrm{for $i \neq j$.}
\end{eqnarray*}
\end{example}

\begin{flushleft}
3.2 Implementation
\end{flushleft}

Let $\{X_{K+1},X_{K+2},\ldots,X_{K+n}\}$ be an irreducible and aperiodic
Markov chain with finite state space $S$ and stationary distribution $\pi$ that satisfies detailed balance.  A burn-in of $K$ draws are discarded, where $K$ depends on the rate of convergence of the sampling algorithm on $\pi$ (Brooks 1998).  We implement our convergence assessment criterion as a test of hypothesis under the null hypothesis that the chain has reached stationarity by iteration $K+1$.  

For $n$ large enough, $V_n \stackrel{\mathcal{D}}{=} \sum_{i=1}^k \lambda_i Z_i^2$, and we estimate its distribution using Lyapunov's Central Limit Theorem (Lo\`eve 1963).  Since $Z_i$ is $\Normal(0,1)$, $Z_i^2$ is $\chi^2_{(1)}$, so $\mathbb{E}\big(\lambda_i Z_i^2 \big) = \lambda_i$ and $\var \big(\lambda_i Z_i^2\big) = 2\lambda_i^2$, for $i=1,\ldots,k$.  Define $Y_i=\lambda_i Z_i^2 - \lambda_i$; $\mathbb{E}(Y_i)=0$, and $\var (Y_i)=\mathbb{E}\big(Y_i^2\big)=2\lambda_i^2 < \infty$ for $i=1,\ldots,n$.  Moreover, $\mathbb{E}\big(Y_i^3\big)=-4\lambda_i^3<\infty$, so $\mathbb{E}\big|Y_i^3\big| < \infty$, for $i=1,\ldots,k$.  Define $s_k^2 = \sum_{i=1}^k \var (Y_i) = 2\sum_{i=1}^k \lambda_i^2$.  It remains to show that the following condition holds:
\begin{math} 
\lim_{k \to \infty} \sum_{i=1}^k \mathbb{E} \big|Y_i\big|^3 / s_k^3= 0,
\end{math}  
which is equivalent to showing that
\begin{equation} \label{CLT:cond}
\lim_{k \to \infty} \frac{1}{\big(2 \sum_{i=1}^k \lambda_i^2 \big)^{3/2}} \sum_{i=1}^k |\lambda_i |^3 = 0,
\end{equation}
since $\mathbb{E} \big|Y_i\big|^3 = |\lambda_i|^3 \mathbb{E} \big|Z_i^2-1\big|^3 \approx 8.6916 |\lambda_i|^3$, for $i=1,\ldots,k$.
So, provided that condition (\ref{CLT:cond}) is satisfied, Lyapunov's Central Limit Theorem gives the following result for $k$ and $n$ large enough:
\begin{equation} \label{CLT:result}
V_n \stackrel{\mathcal{D}}{=} \sum_{i=1}^k \lambda_i Z_i^2 \sim \textrm{Normal} \bigg( \sum_{i=1}^k \lambda_i, 2\sum_{i=1}^k \lambda_i^2 \bigg) \ \textrm{approximately.}
\end{equation}

\noindent
For the computation of the mean and variance in (\ref{CLT:result}), we resort to the following simplifications
\begin{eqnarray} 
\sum_{i=1}^k \lambda_i & = & \textrm{trace}\big(\mathbf{C}\Sigma \mathbf{C}' \big) = \sum_{i=1}^m \big[\mathbf{C}(i,i)\big]^2 \Sigma(i,i), \label{simp1} \\
\sum_{i=1}^k \lambda_i^2 & = & \Big (\sum_{i=1}^k \lambda_i \Big)^2 - 2\sum_{\stackrel{i,j=1}{i<j}}^k \lambda_i \lambda_j, \label{simp2}
\end{eqnarray}
where the first summation in equation~(\ref{simp2}) is given in~(\ref{simp1}), and the second is the sum of all the 2-square principal subdeterminants of $\mathbf{C}\Sigma \mathbf{C}'$ (Marcus and Ming 1964, p. 22).

We propose a quantitative assessment of convergence via a test of hypothesis at confidence level $(1-\alpha)$ using the approximate distribution of $V_n$ given in \eqref{CLT:result} as follows.  
\begin{enumerate}
\item Obtain an aperiodic, irreducible Markov chain which satisfies the principle of detailed balance: 
$\{X_1, X_2, \ldots, X_{K},\ldots, X_{K+n} \}$; discard the first $K$ draws.
\item Compute the statistic $V_n = \frac{n}{m} \sum_{i \in S} \big(f_i - \bar{f} \big)^2$ from the remaining $n$ draws and the $(1-\alpha/2)$ quantile $v_{\alpha/2} = \sum_{i=1}^k \lambda_i + z_{\alpha/2}\sqrt{2 \sum_{i=1}^k \lambda_i^2}$.
\item
If $V_n < v_{\alpha/2}$, conclude that the chain has reached stationarity at level $(1-\alpha)$ and stop; else, continue for an additional $n$ iterations and return to step 2, replacing $n$ by $2n$.
\end{enumerate}

In this article we implement the criterion in the form of a qualitative tool for convergence assessment.  We iterate the chain and plot the absolute value of the relative difference, $\big|\big(V_{(k-1)n}-V_{kn} \big)/V_{(k-1)n}\big |$, against the number of iterations $kn$, every $n$ iterations, $k=1,2,\ldots$.  We claim that the chain has reached equilibrium if the relative difference drops below some problem-specific, pre-specified constant $\epsilon>0$.  The value of the constant $\epsilon$ is problem-specific because it depends on the distribution of interest $\pi$.  For a high-dimensional, multi-modal distribution, the value of $\epsilon$ might need to be very small in order for this analysis to correctly detect lack of convergence to $\pi$, whereas the same value might be too conservative for a one-dimensional, unimodal distribution.  

Based on this implementation of the criterion as a qualitative tool, we can define a measure of efficiency of one algorithm against another.  Let $\epsilon >0$ be given.  Let $V_{n}^{(i)}$ be the value of the statistic after $n$ iterations of algorithm $i, \ i=1,2$.  Let $n_i$ represent the interval, in iterations, at which the statistic is computed for algorithm $i$.  The measure of efficiency is defined as
\begin{displaymath}
V_{1,2}^{(\epsilon)} = \frac{\min \Big \{kn_1: \big| \big(V_{(k-1)n_1}^{(1)} - V_{kn_1}^{(1)} \big )/V_{(k-1)n_1}^{(1)}\big|<\epsilon \Big\}}{\min \Big \{kn_2: \big| \big (V_{(k-1)n_2}^{(2)} - V_{kn_2}^{(2)} \big )/V_{(k-1)n_2}^{(2)}\big|<\epsilon\Big \}}.
\end{displaymath}
If $V_{1,2}^{(\epsilon)} < 1$, we conclude that algorithm 1 is more efficient than algorithm 2 at level $\epsilon$; if $V_{1,2}^{(\epsilon)} > 1$, algorithm 2 is more efficient than algorithm 1.
 
\begin{center}
4. APPLICATIONS
\end{center}
\begin{flushleft}
4.1 Application 1: multipath changepoint problem
\end{flushleft}
The following application is taken from Asgharian and Wolfson (2001).  Let $Y_{ij}$ denote the $j$th measurement on patient $i$, where $1 \leq i \leq 100$, $1 \leq j \leq 20$.  To each patient there is associated a possibly distinct changepoint $\tau_i$ such that measurements $Y_{i1}, Y_{i2}, \ldots, Y_{i\tau_i}$ are i.i.d.\ $\Normal(0,1)$ random variables and measurements $Y_{i\tau_i+1}, \ldots, Y_{i20}$ are i.i.d.\ $\Normal(4,1)$.  Let $Z_i=(1,Z_{i1})'$ and $\theta = (\theta_0, \theta_1)'$ denote the covariate vector and the regression coefficient vector, respectively, for patient $i$, i.e.\ $Y_{ij}=\theta_0 + \theta_1 Z_{i1}, \ \forall j$.  Define parameters $\alpha = \theta_0 + \theta_1$ and $\beta = \theta_0 - \theta_1$.  The goal is to find the maximum likelihood estimators (MLE's) of $\alpha$ and $\beta$, denoted by $\hat{\alpha}$ and $\hat{\beta}$, respectively.  We simulate the data with $\theta_0=0$ and $\theta_1=1$; the joint log likelihood is bimodal.  We let the parameter space be $(-10,10)^2$, assuming zero mass is placed outside this region, and we discretize the space over a grid of width $0.01$.  

We apply the algorithm of simulated annealing, introduced by Kirkpatrick, Gelatt, and Vecchi (1983), which performs function optimization through an iterative improvement approach.  The algorithm was developed via an analogy with thermodynamics where a substance is melted by a slow annealing process and equilibrium is attained at each temperature until eventually the substance stabilizes at its lowest-energy state.  Similarly, in simulated annealing, a global temperature parameter controls the effects of high probability regions under the distribution of interest $\pi$.  For each $T_k$ in a sequence such that $T_k \to 0$ as $k \to \infty$, an MCMC chain with stationary distribution $\pi^{1/T_k}$ is generated until equilibrium.  As the temperature is lowered following a pre-specified schedule, known as the cooling schedule, the effects become more pronounced and the chain stabilizes at its global maximum value or equivalently, lowest energy state (Neal 1993, Brooks and Morgan 1995).  Geman and Geman (1984) show that this convergence is guaranteed under a logarithmic cooling schedule, which unfortunately is too slow to be followed in practice.          

We implement the algorithm with a geometric cooling schedule $T_{k+1} = T_{k}/2, \ k=0,\ldots, 5$, and $T_0=50$ and zero burn-in.  Simulated annealing with a very fast cooling schedule is known as simulated quenching; refer to Catoni (1992) for a discussion on the design of cooling schedules.  For $(\alpha,\beta) \in (-10,10)^2$, the function $f_{\alpha,\beta}^{(k)}$ at temperature $T_k$ is given by
\begin{math}
f_{\alpha,\beta}^{(k)} = \hat{\pi}_{\alpha,\beta} / \exp(-E_{(\alpha,\beta)} ).
\end{math}                      
	  
The aim is to compare the performance of the Metropolis-Hastings sampler in determining the MLE's via simulated annealing with two different methods for proposing the next move.  In the first method, we draw uniformly from a cube of length $w$ centered at the current position, where $w$ has the values: $\{12, 7, 4, 2.5, 1.7, 1.2, 0.9, 0.6\}$ for $k=1,\ldots,8$.  These values are set retrospectively to obtain an acceptance rate of approximately $0.4$. In the second method, we propose the next move via univariate slice sampling applied to each variable in turn; this algorithm is described briefly in Subsection 4.2.   We use the ``stepping-out'' procedure with an initial interval size of $0.1$ at each temperature.  

\begin{figure}
\centerline{\includegraphics[scale=0.4,angle=-90]{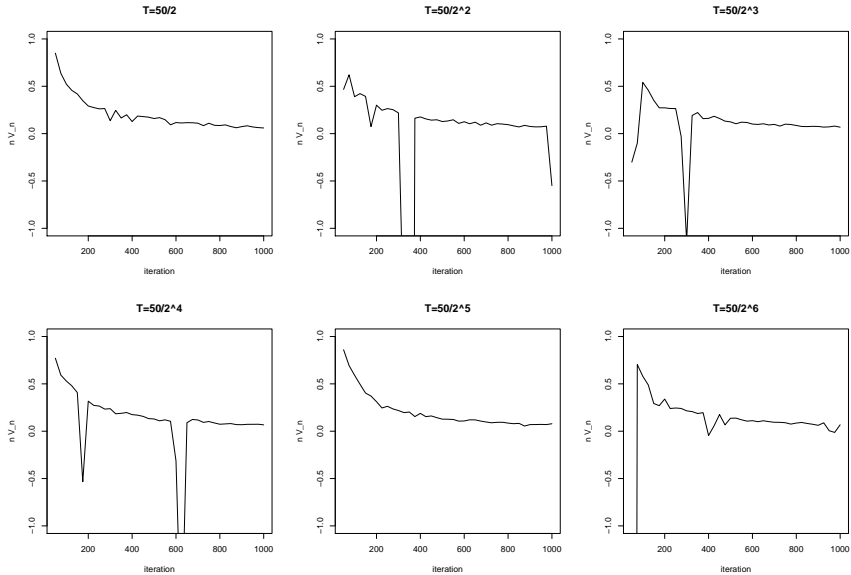}}
\caption{\label{relDiffUnif} Relative difference in $V_n$ versus $n$ using uniform proposal distributions for application 1.  The plots show the decreasing trend of the relative difference in $V_n$ as the number of iterations increases, interrupted by sharp increases in $V_n$.}
\end{figure}

\begin{figure}
\centerline{\includegraphics[scale=0.4,angle=-90]{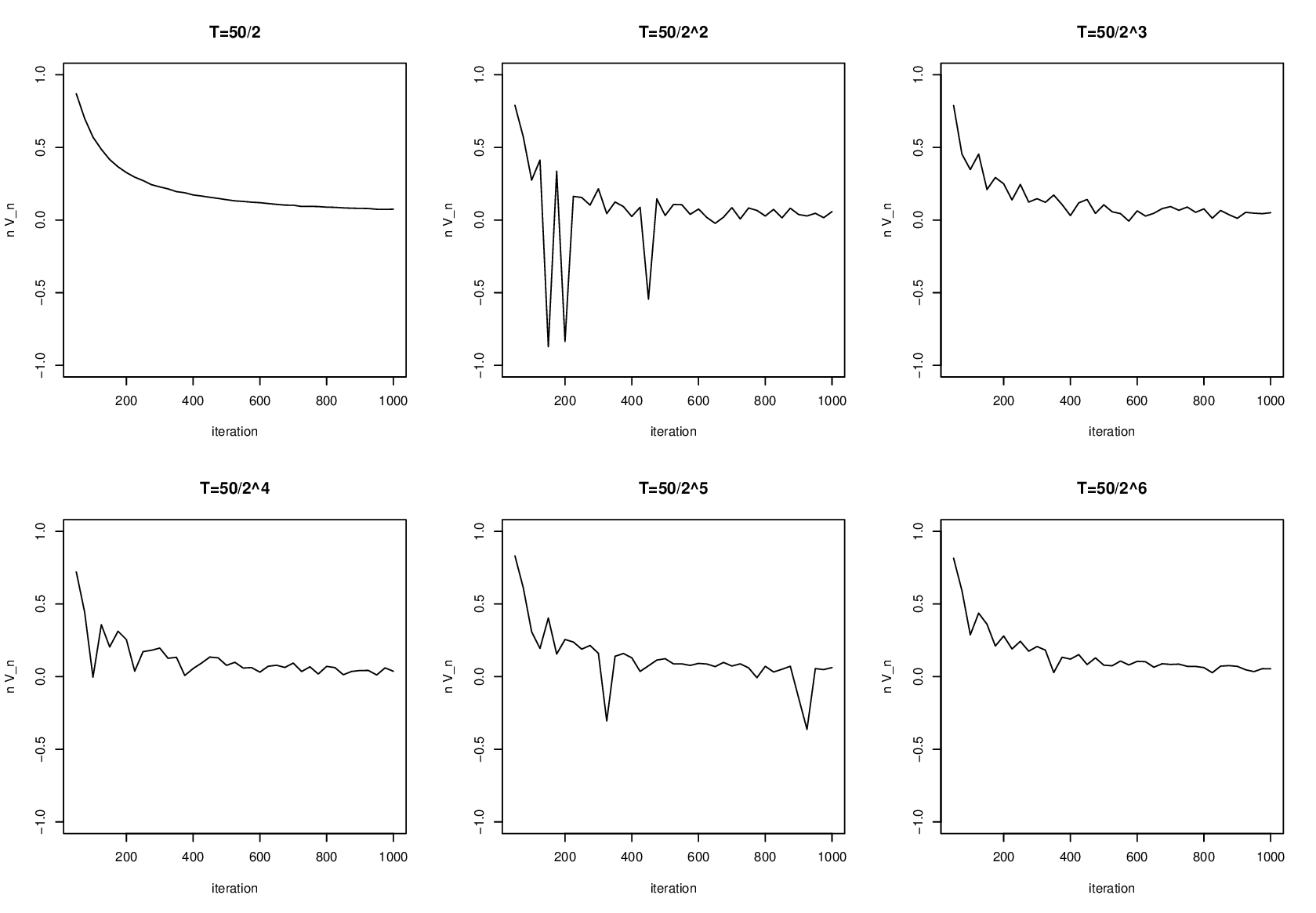}}
\caption{\label{relDiffSlice} Relative difference in $V_n$ versus $n$ using slice sampling for application 1.  The plots show the decreasing trend of the relative difference in $V_n$ as the number of iterations increases; the increases in $V_n$ are more frequent than in Figure~\ref{relDiffUnif}.}
\end{figure}

At each temperature, we perform 1000 iterations of the Metropolis-Hastings algorithm, computing the value of $V_n$ every $25$ iterations.  We obtain the following results: $\big(\hat{\alpha}^{(1)},\hat{\beta}^{(1)}\big)=(1.18,-1.17)$, $E_{(\hat{\alpha}^{(1)},\hat{\beta}^{(1)})}=247.645$, and $\big(\hat{\alpha}^{(2)},\hat{\beta}^{(2)}\big)=(1.19,-1.15)$, $E_{(\hat{\alpha}^{(2)},\hat{\beta}^{(2)})}=247.645$ for the first and second methods, respectively, which equal the lowest energy value obtained by a systematic grid search.  We conclude that both methods correctly identified the MLE's.
Figures~\ref{relDiffUnif} and~\ref{relDiffSlice} display the relative difference in variance; sharp drops indicate that the sampler has jumped to previously unexplored regions of the parameter space, i.e.\ to points $(\alpha,\beta)$ for which $\hat{\pi}_{\alpha,\beta}$ is significantly different from $\pi_{\alpha,\beta}$, thus increasing the value of the variance.

We proceed to simulate 50 datasets; for each, we initialize the two chains from the same randomly chosen point.  At each temperature level, we compute the value of $V_n$ every $25$ iterations until $\big|\big(V_{(k-1)n}-V_{kn} \big)/V_{(k-1)n}\big |<\epsilon$, with $\epsilon=0.05$.  We remark that this value of $\epsilon$ is very conservative; ideally, a different value would be employed at each temperature level.  We make the following two observations: first, for any given dataset, the lowest energy values reported by the two algorithms differ by at most $0.011$ units in magnitude, and, second, the difference between the lowest energy values found by a systematic search and by simulated annealing is at most $0.614909$.  Moreover, we note that the methods required on average 5605 iterations, and 3162 iterations, respectively.  Averaged over 50 tests, the measure of efficiency of simulated annealing using Metropolis-Hastings with uniform proposals versus Metropolis-Hastings with slice sampling is approximately $1.77$, i.e.\ MCMC with slice sampling is almost twice as efficient as MCMC with uniform proposals.

\begin{flushleft}
4.2 Application 2: 10-dimensional funnel
\end{flushleft}
Neal (2003) illustrates the advantage of slice sampling over Metropolis-Hastings in sampling from a 10-dimensional funnel distribution.  Slice sampling is an adaptive MCMC method which proceeds in two alternating steps.  Given the current position $X_t=x_t$, it samples a value $y$ uniformly from the interval $\big (0,\pi(x_t)\big)$.  Given $y$, the next position $X_{t+1}$ is sampled from an appropriately chosen subset of the horizontal ``slice'' $\{x; \pi(x)>y\}$.  Neal (2003) shows that the algorithm produces an ergodic Markov chain with stationary distribution $\pi$, and that, moreover, due to its adaptive nature, the algorithm sometimes outperforms Metropolis-Hastings and the Gibbs sampler.   

Let $X$ be a $\Normal(0,9)$ random variable, and let $Y_1,\ldots,Y_9$ be independent $\Normal$ random variables, which, conditional on $X=x$, have mean 0 and variance $\exp(x)$.  The goal is to obtain an approximate independent sample from the joint distribution of $\big(X, Y_1,\ldots,Y_9 \big)$.  We initialize the chain as follows: $X=0$ and $Y_i=1$, for $i=1,\ldots,9$.  For each variable, the parameter space is taken to be $(-30.0,30.0)$ and it is discretized over a grid of width $0.01$.  

First, we implement the Metropolis-Hastings algorithm with single-variable updates applied to each variable in sequence; one iteration of the chain consists of 1300 updates.  For each variable, the proposal distribution is $\Normal$, centered at the current value, with standard deviation of $1.0$, truncated on the interval $(-30.0,30.0)$.  Numbers are rounded to the closest value on the grid.  Second, we implement the slice sampling algorithm with single-variable updates; each iteration consists of 120 updates for each variable in sequence.  We use the ``stepping-out'' procedure with an initial interval of size $1$.  We compute $V_n$ every $100$ iterations until the absolute value of the relative difference is below $\epsilon=0.01$.

\begin{figure}
\centerline{\includegraphics[scale=0.4,angle=-90]{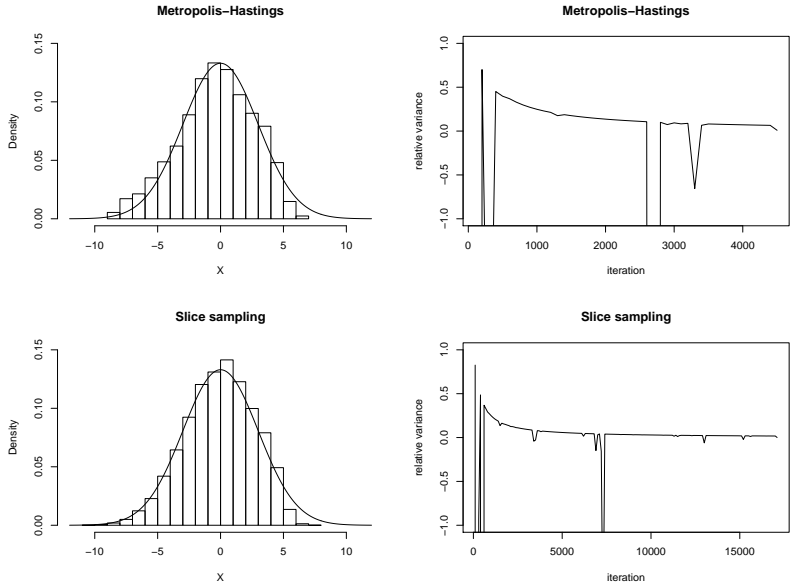}}
\caption{\label{compare_output_v} Sampled values and relative difference in $V_n$ in application 2.  The left column displays histograms of the sampled values of $X$ superimposed on the $\Normal(0,9)$ density function.  The right column displays the relative difference in $V_n$ versus $n$.}
\end{figure}

\begin{figure}
\centerline{\includegraphics[scale=0.4,angle=-90]{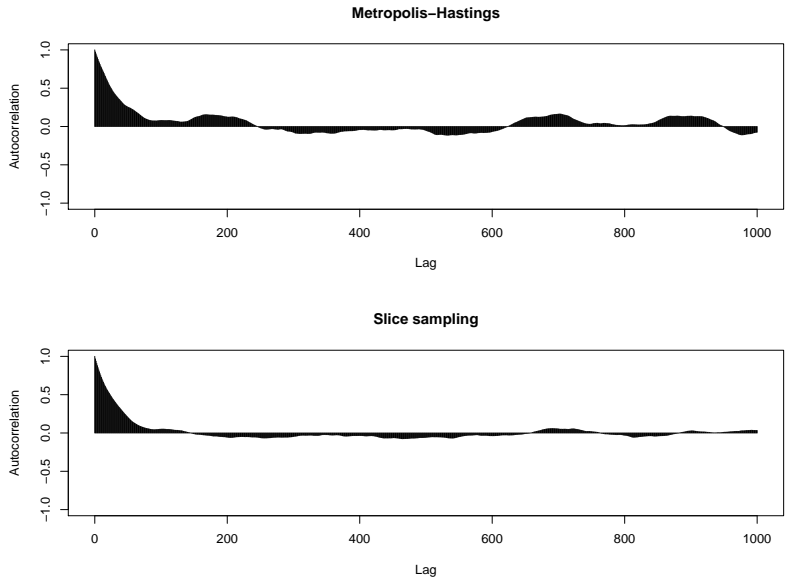}}
\caption{\label{autocorrelations} Autocorrelation of $X$ in application 2.  Slice sampling has a faster rate of convergence than Metropolis-Hastings evidenced by the smaller autocorrelation.}
\end{figure}

\begin{figure}
\centerline{\includegraphics[scale=0.4,angle=-90]{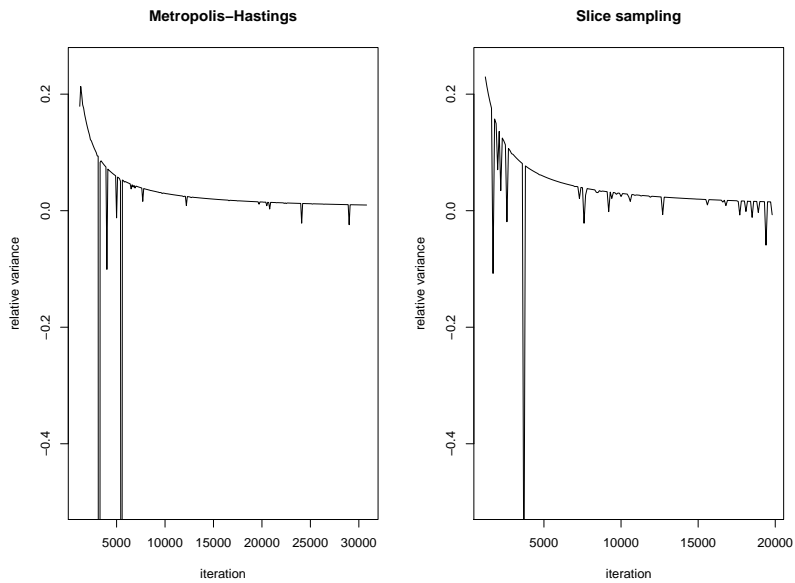}}
\caption{\label{relDiff_10dim_v_parallel} Relative difference in $V_n$ versus $n$ for eleven parallel chains in application 2.  The value of $V_n$ under Metropolis-Hastings sampling seems to be more stable than under slice sampling.}
\end{figure} 

The left column of Figure~\ref{compare_output_v} compares the histograms of the sampled values of $X$ with the true probability distribution function; the histograms are based on chains of 4600 and 17200 iterations, respectively.  Metropolis-Hastings oversamples negative values of $X$ and undersamples positive ones; slice sampling samples correctly in the left tail of the distribution, but undersamples positive values.  The right column displays the behaviour of the relative difference in $V_n$; the variance function undergoes sharp increases in value under both sampling methods, but stabilizes towards the end of the run.  The behaviour of the variance function fails to reflect the incorrect sampling in the tails of the distribution.  The plot of the relative difference in variance for the Metropolis-Hastings algorithm indicates that a smaller value of $\epsilon$ would be more appropriate for assessing convergence. The plots in Figure~\ref{autocorrelations} show that the autocorrelation obtained by slice sampling remains close to zero after 100 iterations, whereas that obtained by Metropolis-Hastings continues to fluctuate even after 1000 iterations.  This indicates that the Metropolis-Hastings algorithm converges more slowly than slice sampling.  We compute the Raftery and Lewis (1992) convergence diagnostic using the Coda package in R (http://www.r-project.org) obtaining dependence factors of 14 and 18.7 for the Metropolis-Hastings and the slice sampling algorithms, respectively, indicating strong autocorrelation.

Finally, we run eleven parallel chains started from the following quantiles of the marginal distribution of $X: \{0.1, 0.2, 0.3, 0.4, 0.45, 0.5, 0.55, 0.6, 0.7, 0.8, 0.9\}$; we employ the value $\epsilon=0.01$.  We expect the parameter space to be insufficiently explored by both algorithms;  however, we are interested in whether this insufficient exploration can be detected from the behaviour of $V_n$ across chains with overdispersed starting points.  Pooling the sampled values results in chains of 30800 and 19800 draws, respectively; thus the measure of efficiency of Metropolis-Hastings versus slice sampling is 1.56.  Trace plots and histograms indicate that negative values of $X$ are oversampled and positive ones are undersampled by both algorithms.  Figure~\ref{relDiff_10dim_v_parallel} is obtained by pooling the sampled values across the eleven chains; the behaviour of $V_n$ under slice sampling poses signs of concern regarding convergence to stationarity (notice the frequent increases in value from iteration 17500 onwards), whereas the value of $V_n$ under Metropolis-Hastings appears stable towards the end of the run.  Therefore the behaviour of $V_n$ under slice sampling across eleven chains with overdispersed starting points indicates lack of convergence to stationarity, whereas the behaviour of $V_n$ under Metropolis-Hastings, which is known to allow a more restrictive exploration of the support space, gives misleading results.
  
\begin{center}
5. CONCLUSION
\end{center}
The last fifty years have witnessed the development and rise in popularity, in particular in Bayesian statistical inference, of Markov Chain Monte Carlo methods for simulating from complex probability distributions (Smith and Roberts 1993).  For a practitioner who has a finite MCMC output, questions arise regarding how reliable the sample is as a representation of $\pi$.  Although a wealth of convergence diagnostic tools for analysing MCMC output have been proposed over the past decades, their performance, in general, is problem-specific, and developing a dependable, easy to implement tool for convergence assessment continues to be a challenge.  This article presents a new convergence assessment method for irreducible, aperiodic Markov chains on discrete spaces obtained by MCMC samplers that satisfy the principle of detailed balance and requirement (\ref{CLT:cond}).  We introduce a one-dimensional test statistic whose behaviour under the assumption of stationarity is analyzed both theoretically and experimentally, and present a possible implementation of our criterion as a graphical tool for convergence assessment.  

In low dimensional problems, the proposed criterion as a qualitative tool assesses convergence satisfactorily; however, in high dimensional problems, the criterion is unreliable for convergence assessment, but can provide useful insight into lack of convergence of the chain to stationarity.  In particular, if the variance function experiences sharp increases in value, then it can be concluded that stationarity has not yet been reached; however, if the value of the variance function is stable, then the results are inconclusive.  The advantage of our method lies in its attempt to analyse the behaviour of an MCMC chain travelling through a possibly high dimensional space by monitoring the behaviour of a one-dimensional statistic.  Lack of convergence to stationarity is correctly assessed by the behaviour of the statistic to the extent to which the sampler explores freely the underlying space.  Particularly in high dimensional problems with irregularly shaped distribution functions, we recommend that the MCMC output be analyzed using different $\epsilon$ values, compared across multiple chains, and that several diagnostic tools be employed.  

There exist in the literature at least two convergence assessment criteria based on weighting functions that are very similar to our approach.  Ritter and Tanner (1992) propose to detect convergence to the full joint distribution by monitoring convergence of the importance weight $w_t=\pi(x)/g_t(x)$, where $g_t$ is the joint distribution of the observations sampled at iteration $t$.  They estimate $g_t(x)$ by $\frac{1}{m} \sum_{i=1}^m p\big(x | x_{t-1}^{(i)}\big)$, where $x_{t-1}^{(i)}, \ i=1,\ldots,m$ is a sample from $g_{t-1}$.  If the chain has converged, the distribution of the weights $w_t$, based on multiple replications of the chain, will be degenerate about a constant.  Zellner and Min (1995) propose a convergence criterion for the Gibbs sampler in the special case that $x$ can be partitioned into $\big(x_{(1)},x_{(2)}\big)$.  They define two criteria based on the weight functions $W_1=p(x_{(1)})p(x_{(2)} | x_{(1)}) - p(x_{(2)})p(x_{(1)} | x_{(2)})$ and $W_2= \big [p(x_{(1)})p(x_{(2)} | x_{(1)})\big]/\big [p(x_{(2)})p(x_{(1)} | x_{(2)}) \big]$, where $p_{(1)}$ is estimated by $\frac{1}{m} \sum_{i=1}^m p\big (x_{(1)} | x_{(2)}^j \big)$, and $x_{(2)}^j, \ j=1,\ldots,m$ is the sequence of draws of $x_{(2)}$ obtained by Gibbs sampling.  They compute the value of these weights at many points in the parameter space and argue that if the chain has converged, then the values of $W_1$ will be close to 0 and those of $W_2$ close to 1.  Zellner and Min use asymptotic results from the stationary time series literature to calculate posterior odds for the hypothesis $H_0: W_1=0 \ $ vs. $\ H_1: W_1 \neq 0$ for the $k$-dimensional case, $k \geq 1$, when the weights are computed at $k$ different points in the parameter space. 

The main drawback of these methods is the assumption that the transition probability $p(x | x_{t-1})$, in the method of Ritter and Tanner, and the conditionals $p(x_{(1)} | x_{(2)})$ and $p(x_{(2)} | x_{(1)})$, in the method of Zellner and Min, exist explicitly.  Our method, however, makes no such assumption and estimates $\pi_i$, the probability of being in state $i$, by the empirical distribution function.  All three methods have the disadvantage of being computationally expensive; the ergodic averages used to approximate various marginal and conditional probabilities (in our method, $\hat{\pi}_i$) require a large number of summands in order to provide good estimates, so large numbers of iterations, and possibly many replicates of the chain, are needed.  Furthermore, since the normalizing constant of $\pi$ is unknown, the functions $f_i$ and the weights $w_t$ of the criterion of Ritter and Tanner might stabilize around an incorrect value if the sampler has failed to explore all the high density regions of the space.  For this reason, we recommend to run multiple replicates of the chain started from different regions of the space.  The criterion of Zellner and Min also gives misleading results if the space is poorly explored and the weights are computed at points that come from low density regions.  Finally, our criterion has an intuitive graphical representation, very similar to that proposed by Ritter and Tanner, and, whereas the criterion of Zellner and Min uses multivariate weight functions, our criterion is based on a one-dimensional statistic regardless of the dimension of the underlying space, thus offering a dimensionality reduction approach to the problem of convergence assessment in high dimensional spaces.  

An interesting alternative to approximating a continuous state space by a discrete grid is to sample the continuous state-space Markov chain and to apply the discretization method developed by Guihenneuc-Jouyaux and Robert (1998).  Provided that the continous chain is Harris-recurrent, the method defines renewal times based on the visiting times to one of $m$ disjoint small sets in the support space.  By subsampling the underlying chain at the renewal times, the method builds a homogeneous Markov chain on the finite state space $\{1,\ldots,m\}$.  Our propoposed criterion can then be applied to the finite chain; it would be interesting to explore whether the convergence assessment extends to the continous Markov chain.

\end{document}